\documentclass{article}
\usepackage{graphicx}
\def\p{\partial}

\def\g{\gamma}

\def\de{\delta}

\def\De{\Delta}
\def\ov{\overline}

\def\ld{\lambda}
\def\Ld{\Lambda}

\def\th{\theta}

\def\Om{\Omega}
\def\rh{\rho}

\def\pdellx'{\frac{\partial}{\partial x'}}
\def\pdellw'{\frac{\partial}{\partial w'}}

\newcommand{\be}{\begin{equation}}
\newcommand{\ee}{\end{equation}}
\def\bed{\begin{displaymath}}
\def\eed{\end{displaymath}}
\def\bea{\begin{eqnarray}}
\def\eea{\end{eqncrray}}
\def\[{$$}
\def\]{$$}
\begin{document}
\UseRawInputEncoding
 \title{A Model for Dark Energy Based on Baryonic Charges with General Yang-Mills Symmetry}
\author{
Jong-Ping Hsu\\
 Department of Physics, \\
 University of Massachusetts Dartmouth,\\ 
North Dartmouth, MA 02747, USA\\
Leonardo Hsu\\
Department of Chemistry and Physics,  \\   Santa Rosa Junior College,\\
Santa Rosa, CA 95401, USA\\}

\maketitle
{\small  General Yang-Mills symmetry and Yang-Mills gravity in inertial frames presents a possible avenue for understanding dark energy phenomenon in terms of the extremely small baryon charges of protons and neutrons.  General gauge transformations involve  vector gauge functions and Hamilton's characteristic phase functions rather than scalar gauge functions and usual phase functions. One has fourth-order gauge field equations, which lead to a static linear potential between two point baryonic charges. Together with gravitational force, the model predicts  that in the early universe, when the gravitational force was dominant, not only was the expansion occurring at a decreasing rate, the rate of change of the deceleration was decreasing  due to a $1/r^2$ dependence. In the present universe, when the baryonic force is dominant, the cosmic acceleration should be approximately constant and has a maximum value $a_{max}\approx 10^{-13} m/s^2$, as measured in an inertial frame.  We hope that these properties of the late-time cosmic acceleration can be tested in the future.}




\section{Introduction}

The phenomenon of the late-time accelerated cosmic expansion is so far from our expectations that it seems very difficult to devise a reasonable explanation for it based on established physics.  However, Broader Particle-Cosmology with general Yang-Mills symmetry presents a possible avenue for understanding this phenomenon in terms of the baryonic charges of protons and neutrons.  Conventionally, one would consider the baryon charge to be associated with the conventional $U_1$ gauge symmetry, just like the leptonic or electric charges. However, Broader Particle-Cosmology allows the possibility that the baryon charge could instead be associated with a general Yang-Mills $U_1$ symmetry, whose gauge transformations involve an arbitrary vector gauge function rather than the conventional scalar gauge function.\cite{1} Although the conservation of baryon charge has been established experimentally, its corresponding gauge field  $B_\mu$ has not been detected in high energy laboratories.  This may be due to the extreme smallness of the baryonic charges to be detected in Earth-bound laboratory.  

The key implication of associating baryonic charge with a general Yang-Mills $U_1$ symmetry is that one has a new gauge invariant field equation, which leads to a static linear potential and thus a distance-independent force between two point baryonic charges.\cite{1} This force between two baryons would be repulsive, while the force between a baryon and an anti-baryon would be attractive. The existence of such a force makes possible a situation where billions of years ago, when the galaxies were closer together,  the attractive gravitational force was dominant, leading to a cosmic expansion taking place at a decreasing rate. However, when the distance between galaxies exceeded a certain critical distance, the repulsive baryon-baryon force became dominant, resulting in the accelerated expansion we now observe. Naturally, it must be the case that the baryon-baryon force is extremely small so that it has thus far evaded detection in the lab, becoming important only when cosmic-sized objects with enormous numbers of baryons, are involved. 
  
 Although the general Yang-Mills $U_{1}$ symmetry involves vector gauge functions $\Ld^\mu(x)$ and leads to fourth-order field equation, the conventional $U_1$ gauge symmetry is merely a special case of the general Yang-Mills $U_{1}$ symmetry where the vector gauge function $\Ld_\mu(x)$ can be expressed in terms of a  scalar gauge function $\Ld(x)$, where $\Ld_\mu(x)=\p_\mu \Ld(x)$. As a result, the usual gauge field equation is a second-order differential equation.  This situation where the usual gauge symmetry is a special case of a general Yang-Mills symmetry also holds for other symmetries such as $SU_2$ and $SU_3$.\cite{1,2}

\section{Dynamics of baryonic 
charges }
Since the baryonic dynamics with general Yang-Mills symmetry is new, let us briefly review the baryonic Lagrangian with general $U_1$ gauge symmetry, which is given by 
\be
L_{B}= + i\ov{q} \g^\mu (\p_\mu + ig_b B_\mu)q
- m_q \ov{q} q +\frac{L_b^2}{2}\p^\mu B_{\mu\ld} \p_{\nu}B^{\nu\ld} 
\ee
where $c=\hbar=1$ and  $g_b$ is the baryon charge of a single quark (an anti-quark would have a charge of -$g_b$).   The summation in the quark terms, such as $m_q \ov{q}q$, is understood.\cite{2}  Since the matter universe is dominated by protons and neutrons, we may just consider up quark and down quark in the baryonic Lagrangian $L_B$ in (1).

The gauge curvature $B_{\mu\nu}$ associated with $B_\mu(x)$ is defined as usual by the commutators
\be
[\De_{B\mu} ,\De_{B\nu}]=  ig_{b} B_{\mu\nu},  \ \ \  B_{\mu\nu}=\p_\mu B_\nu -\p_\nu B_\mu, 
\ee
where $\De_{B\mu}$ is the usual gauge covariant derivatives
\be
\De_{B\mu}=\p_\mu +ig_b B_\mu(x), \ \ \ \  c=\hbar=1. 
\ee

For baryons, we assume the general Yang-Mills $U_{1b}$ symmetry, whose gauge transformations involve general vector gauge functions\footnote{In general, the vector gauge function $\Ld_\ld(x)$ cannot be expressed in terms of a scalar function.} $\Ld_\ld(x)$ and Hamilton's characteristic phase factor $ \Om_{B}(x)$,\cite{3,4}
 \be
q'(x) =   \Om_B(x) q(x), \ \ \ \ \ \  \ov{q}'(x) = \ov{q}(x)  \Om^{-1}_{B}(x),
\ee
\be
 \Om_{B}(x) = e^{-iP(x)}, \ \ \   
 \ee
 \be
 P(x)=\left(g_b \int_{x'_o}^{x'_e=x} dx'^\ld \Ld_\ld (x') \right)_{Leb} ,
\ee
\be
\de P(x)=g_b \Ld_\mu(x') \de x'^\mu|_{x_o'} ^{x} 
\ee
$$
- g_b\int^{x}_{x'_o} [\p'_\mu \Ld_\ld(x')-\p'_\ld \Ld_\mu(x')]dx'^\mu \de x'^\ld. 
$$
 The subscript `$Leb$' in (6) denotes that the path of the action integration in     $\de P(x)$ must satisfy the equation,
\be
 [\p_\mu \Ld_\ld(x)-\p_\ld \Ld_\mu(x)]dx^\mu=0.
\ee
Thus, Hamilton's characteristic phase factor $\Om_B(x)$ in (5) is unambiguously  a local function of $x$.\cite{3,4}  The variation $\de P(x)$ in (5)-(7) and the constraint (8) leads to the local relation at any space-time point $x$,
\be
\frac{\p P(x)}{\p x^\mu} = g_b \Ld_\mu(x),
\ee
which is an important relation for the general $U_{1b}$ symmetry.  

The coupling between the baryonic `phase field' $B_\mu(x)$ and the quark field $q(x)$ in (1) are invariant under new general $U_1$ gauge transformations involving vector gauge functions $\Ld_\mu(x)$:
\be
\p^\mu B'_{\mu\nu}(x)= \p^\mu B_{\mu\nu}(x), \ \ B'_\mu(x)= B_\mu(x) + \Ld_\mu(x).
\ee
\be
   \p^\mu\p_\mu \Ld_\nu (x) - \p^\mu\p_\nu \Ld_\mu (x) = 0;
\ee
\be
\ov{q}'(x)\g^\mu \De'_{B\mu} q' (x) =\ov{q}(x)\g^\mu\De_{B\mu} q (x), \ \ \ \ \    
\ee
 One can verify that the usual gauge curvature $B_{\mu\nu}$ is, in general, no longer invariant under the new general gauge transformations involving the vector gauge function $\Ld_\mu$.  However, the divergence  of the gauge curvature, i.e., $\p^\mu B_{\mu\nu}$, in (10) can be invariant under the new gauge transformations, provided the constraint (11) is imposed.  This constraint is automatically satisfied if $\Ld_\mu(x) =\p_\mu \Ld(x)$, which does not hold in the general $U_1$ gauge symmetry.  The local relation (9) is crucial for the general Yang-Mills invariant coupling of quarks in (12).  

 Thus, the new transformations in (4)-(12) define the  general $U_{1}$ gauge transformations for baryonic dynamics based on the Lagrangian (1). The corresponding vector fields $B_\mu$  are called `phase fields.' 
  
\section{The distance-independent baryon-baryon force for two point-like baryonic charges}

The fourth-order equations\cite{5,6} of the phase field $B_\mu$ can be derived from the Lagrangian (1), 
\be
L_{b}^2 \p^2 \p^\ld B_{\ld\mu} -  g_b\ov{q} \g_\mu q = 0,
\ee
where $L_b$ is a constant with the dimensions of length (i.e., a `fundamental' length).
 We impose the gauge condition $\p^\ld B_\ld =0$ and consider a single quark at the origin.  The zeroth component static gauge potential $B_0({\bf r})$ satisfies 
\be
\nabla^2 \nabla^2 B_0 ({\bf r})=  \frac{g_b}{L^2_b} \de^3({\bf r}).
\ee
The potential $B_0({\bf r})$\cite{5,6} results in a repulsive force $F_{qq}$  between two point-like quarks,
\be
B_0({\bf r}) =- \frac{g_b}{L^2_b}\frac{ r}{8\pi}, \ \ \ \   {\bf F}_{qq}=- g_b 
{\nabla} B_0=\frac{1}{8\pi}\frac{g^2_b}{L^2_b}\frac{{\bf r}}{r},
\ee
where we have used the following relation for generalized functions,\cite{5}   
$$
\int^{\infty}_{-\infty}{1}/{({\bf k}^2)^2} exp(i{\bf k .  r}) d^3 k = -\pi^{2} r.
$$
This force is independent of the distance between the two quarks. It is attractive between a quark and an anti-quark, and repulsive between two quarks or two anti-quarks.  The repulsive force between two extended baryonic galaxies is also approximately independent of their distances,\cite{7} as shown in Appendix. This force was called the cosmic Okubo force.\cite{1,8} 

\section{Okubo forces on a cosmic scale}

Although the values of $g_b$ and $L_b$ in (1) are unknown, we can estimate them as follows. There is data that suggests that rate of expansion of the universe was not always increasing and that the accelerated expansion began only about 4 billion years ago, when the universe was approximately 9 billion years old.\cite{9} From this information, one could estimate the size of the universe at this critical point to be $V_{uc} \approx R_{uc}^3 \approx  (9 \ billion \ years \times speed \ of \ light)^3 \approx 6 \times 10^{77} m^3$.

For simplicity, we consider two galaxies with equal masses $m_g$. Because the vast majority of the mass of a galaxy consists of hydrogen, virtually all of a galaxy's mass is baryonic (protons, in this case). The baryonic charge of each galaxy may thus be approximated as $(3g_b)m_g/m_p$, where $m_p$ is the proton mass and $g_b$ is the baryonic charge of a single quark. The critical distance $d_c$ between two galaxies, i.e., the inter-galactic distance beyond which the repulsive baryon-baryon force becomes larger than the attractive gravitational force, can be found by setting the magnitudes of the gravitational and Okubo forces equal

\be
| {F_{gra}}|=\frac{Gm_g^2}{{d_c}^2} \approx  {F_{Ok}}\approx \frac{(9 g_b^2 m_g^2 )}{(8\pi L_b^2 m_p^2 )},  
\ee
\be
d_c \approx \left(\frac{G m_p^2 8 \pi L_b^2 }{ (3g_b)^2} \right)^{1/2}, 
\ee 
where we have treated the galaxies as point masses because the distance between galaxies is typically much larger than the size of a galaxy. In general, when a galaxy is modeled as a sphere, the force between two galaxies is also approximately independent of distance.  (See Appendix.)  We assume that the galaxies in the universe are uniformly distributed (consistent with the conventional cosmological principle, which approximates the universe as homogeneous and isotropic). When the average distance between two galaxies is $d_c$, the volume of space per galaxy is approximately $d_c^3$.  The total volume $V_{uc}$ of the universe is approximately 
\be
V_{uc}\approx  N_g d_c^3 \approx N_g \left(\frac{\sqrt{8\pi G} L_b m_p}{3 g_b} \right)^3,
\ee
where $N_g$ is the number of galaxies in the universe. Current estimates for the number of galaxies in the universe $N_g$ vary, but a representative number is $N_g\approx 2 \ trillions.$\cite{10}  Setting this equal to our estimate for the size of the universe 4 billion years ago, we estimate that
\be
\ \ \ \  g_b/L_b  \approx 3\times10^{-42}/m, \ \ 
\ee
where we have used $ d_c \approx 10^{22} m$.

To find values for $L_b$ and $g_b$ individually, we note that the length $L_b$ appears as a fundamental length in the Lagrangian (1) for baryonic dynamics.  There is also a fundamental length $L_s=0.28 fm$\cite{11} in the Lagrangian for the charmonium associated with the general Yang-Mills $SU_3$ symmetry.  From the viewpoint of a total unified model, including baryonic and color charges, it is perhaps reasonable to assume that the Lagrangian of the general Yang-Mills symmetry\cite{1} has only a single fundamental length, i.e.,
\be
L_b = L_s \approx 0.28\times 10^{-15} m.
\ee  
This gives a value for the dimensionless baryonic charge $g_b$ of
\be
g_b\approx 2\times 10^{-57}.      
\ee

This value is an unimaginably  small coupling constant in particle physics and the Okubo force can probably never be detected in Earth-bound laboratories.  Nevertheless, it may have observable cosmological effects.  This result suggests an interesting relation based on the principle of gauge invariance that a microscopic physical quantity such as a basic length in quark confinement $L_s=0.28 fm$ could be related to the baryon charge $g_b\approx 2 \times 10^{-57}$ in (21) for the super-macroscopic phenomenon of cosmic acceleration.  This would provide a coherent big picture of unification as suggested in the Broader Particle-Cosmology.\cite{12,13}

\section{Summary of the model of dark energy}
We summarize the features of the proposed model of dark energy as follows:

(A)  The proposed model is based on Broader Particle-Cosmology with general Yang-Mills symmetry for conserved baryon charges.  In particular, the basic principle of gauge symmetry in inertial frames is rooted in established physics rather than ad hoc assumptions.  Conventional models of dark energy assert that the universe is composed of 4\% baryons, 20\% dark matter, and 76\% dark energy.\cite{9}  Based on Broader Particle-Cosmology,\cite{13} we have demonstrated that dark matter phenomena could be due to the enhanced Yang-Mills gravity of  high energy anti-e-neutrinos. In the previous discussions, we have demonstrated that the dark energy phenomena could be generated by a cosmic Okubo force associated with well-established baryonic charges.   Together, this eliminates the need to postulate the existence of heretofore unknown forms of matter or energy that make up 96\% of the cosmos. As a bonus, it is gratifying that Broader Particle-Cosmology is based on flat space-time, in which the space-time coordinates in inertial frames have well-established, operational definitions, in contrast to the conventional models of dark energy based on general relativity, with its more complicated curved space time that is incompatible with quantum field theories.

(B)  Like the Coulomb and gravitational forces, the effects of the repulsive Okubo force should be felt  instantaneously by objects at all distances\cite{14,15}  Consequently, the dynamics of objects separated by enormous distances can be calculated similar to phenomena in laboratories, without needing to account for a propagation time for those forces. 

(C)  The model based on the cosmic Okubo force of baryon charges predicts the late-time cosmic acceleration to be 
\be
a_{cos}(r)=\frac{ (F_{Ok} + F_{gra})}{m_g }=\frac{9 m_g g^2_b}{8\pi m^2_p L_b^2}\left(1-\frac{d^2_c}{r^2}\right),
\ee
\be
\frac{9 m_g g^2_b}{8\pi m^2_p L_b^2}= a^{max}_{cos}\approx 10^{-13} \frac{m}{s^2}, \ \ \ \  (SI \ units)
\ee
where we have used (16) and r is the average distance between two galaxies.  All numerical calculations are order of magnitude estimations, and carried out in inertial frame based on the framework of Broader Particle-Cosmology.  It appears that there is no simple way to compare the maximum cosmic acceleration $a^{max}_{cos}$ in (23) with the usual experimental data\cite{9} for cosmic expansion expressed in the units Km/(s MegaParsec) based on a different conceptual framework in curved space-time.  Moreover, there are other explanations for the origins of cosmic acceleration based on, say, a non-perturbative Yang-Mills condensate.\cite{16}. They are conceptually different from our model and, hence, it is difficult to compare such a model with our model, which gives specific results (22) and (23).

(D)  According to the proposed model, the acceleration of a galaxy is the combined result of the repulsive Okubo force and the attractive gravitational force on it by other galaxies. Thus, a main prediction of the model is that in the early universe, when the gravitational force was dominant, not only was the expansion occurring at a decreasing rate (decelerating), the rate of change of the deceleration was decreasing (since the gravitational force has a $1/r^2$ dependence. In the present universe, when the Okubo force is dominant (as evidenced by the accelerating expansion), the acceleration should be constant approximately, as shown in (22) and (23).  All numerical results in this work are order-of-magnitude estimations and all physical quantities are measured in inertial frames, as required by Broader Particle-Cosmology with Yang-Mills gravity and general gauge symmetry.  Cosmic redshifts  based on Yang-Mills gravity and recession velocity have been discussed in inertial frames and constant-acceleration frames.\cite{17}  We hope that these properties of the late-time cosmic acceleration can be tested in the future.

The work was supported in part by Jing Shin Research Fund and Prof. Leung Memorial Fund of the UMassD Foundation.

\bigskip
\newpage
\noindent
{\large \bf   Appendix.   \  Approximately distance-independent 
 Okubo force between two galaxies } 
\bigskip

Let us consider the case that a galaxy S, which is located outside another galaxy modeled as a sphere with uniform baryon charges and a radius $R_o$. Our calculation shows that the Okubo force between two galaxies is approximately distance-independent.  

\begin{center} 
\hspace*{-2cm}      
\includegraphics[width=8.cm]{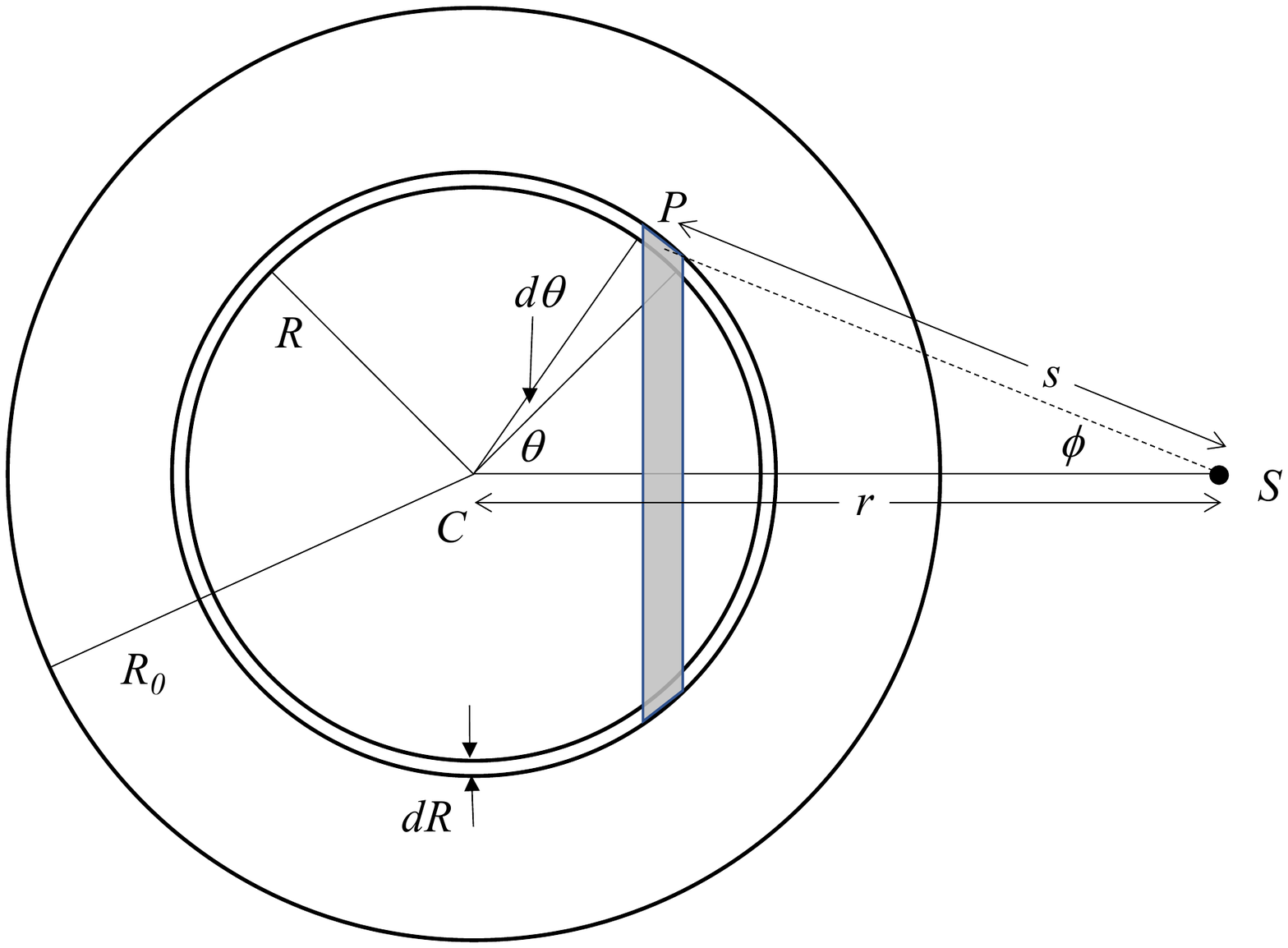}
\end{center}
\vspace*{-0.4 cm}
Fig. 1.  A schematic diagram for calculations of the baryonic Okubo force between two uniform spheres of baryonic charges.  One sphere is considered to have all its baryonic charges concentrated at its center S.

\bigskip

To calculate the force on $S$ by the uniform sphere, we first divide the sphere into thin spherical shells with a radius $R<R_o$ and thickness $dR$.  Now consider dividing each spherical shell into rings, as indicated by the shaded region in Figure 1.  The mass $dM_r$ of the ring is $dM_r = \rh (2\pi R  sin\th ) (R d\th)(dR),$ 
where $2\pi R sin\th$ 
is the circumference of the ring, $R d\th$ is the width of the ring, and $dR$ is the thickness of the ring. Now let the distance between $S$ and every point of the ring be $s$ and the angle PSC to be $\phi$, as shown in Fig. 1. By symmetry, the baryonic force $dF_{CS}$ on $S$ by the ring must point along CS and, using (3.21), is given by 
\renewcommand\theequation{{A1.1}} 
\be
dF_{CS}=\int\frac{1}{8\pi L^2_b}\frac{3g_{b} m}{m_p}\frac{3g_b dM_r}{m_p} cos\phi ,   \ \ \ \ \      
\ee
where
\renewcommand\theequation{{A1.2}} 
\be
 dM_r=(\rh 2\pi R^2 sin\th d\th) dR.  
\ee
In (4.1), $3g_b m/m_p$ is the baryonic charge of $S$, $3g_b dM_r/m_p$ is the baryonic charge of the ring, and $cos\phi$ results from taking only the component of the force on $S$ by the ring that is parallel to CS (by symmetry). 

To evaluate this integral, we note that for the triangle CSP, we have 
\renewcommand\theequation{{A1.3}} 
\be
s^2=R^2+r^2 -2Rr \ cos\th, \ \  \ \   R^2=s^2 +r^2 -2rs \ cos\phi
\ee
 by the law of cosines.  The integration of $d\th$ from $0$ to $\pi$ leads to an effective repulsive force on $S$ along CS.  
 The magnitude of the total repulsive force  can be obtained\cite{7}ok ok from the integration of $ds$:
 \renewcommand\theequation{{A1.4}} 
\be
dF_{CS} =A\int _{0}^{\pi} \left[dR \ R^2 cos\phi\right] sin\th \ d\th = A \int_{r-R}^{r+R} [dF ]\frac{s}{Rr}ds,
\ee
\renewcommand\theequation{{A1.5}} 
\be
 A= \frac{(3g_b)^2 M \rh}{4 L^2_b m_p^2},  \ \ \  dF \equiv dR \ R^2\left( \frac{s^2+r^2-R^2}{2sr}\right).
\ee

Thus, the total repulsive baryonic force $F^{Ok}$ exerted on the baryonic charges $+3g_{b}M/m_p$ in the 
outside of the gigantic sphere 
with baryonic charges $3g_{b}M/m_p$ is 
\renewcommand\theequation{{A1.6}} 
\be
F^{Ok}=\int_0^{R_o} \frac{dF_{CS}}{dR} \ dR = \left(\frac{(3g_b)^2 M^2}{8\pi L^2_b m^2_p}\right)\left[1-\frac{R^2_o}{5r^2}\right], 
\ee
\renewcommand\theequation{{A1.7}} 
\be
\approx  \left(\frac{(3g_b)^2 M^2}{8\pi L^2_b m^2_p}\right),  \ \ \ \ \ \  R_o << r,  \ \ \ \ M=\frac{4\pi R_o^3\rh}{3}.
\ee
 When $r^2>> R^2_o$, the cosmic Okubo force $F^{Ok}$ is approximately a constant, i.e., independent of the distance $r$.
 
 \bigskip

\bigskip 
\bibliographystyle{unsrt}

\end{document}